\documentclass[]{elsart}
\usepackage{latexsym}
\usepackage[final]{graphicx}

\usepackage{natbib}
\usepackage{amssymb}

\newcommand{\eg}{e.g.}
\newcommand{\msun}{\ensuremath{M_{\odot}}}
\newcommand{\alp}{\ensuremath{\alpha}}

\newcommand{\gcc}{\ensuremath{\mathrm{\thinspace g \thinspace cm^{-3}}}}
\newcommand{\ye}{\ensuremath{\mathrm {Y}_{e}}}

\begin{document}

\begin{frontmatter}
\title{Nucleosynthesis in Neutrino-Driven Supernovae}

\author[AUT1]{C. Fr\"ohlich}
\author[AUT2]{W. R. Hix\thanksref{coraut}}
\author[AUT3]{G. Mart{\' i}nez-Pinedo}
\author[AUT1]{M. Liebend\"orfer}
\author[AUT1]{F.-K. Thielemann}
\author[AUT4]{E. Bravo}
\author[AUT3]{K. Langanke}
\author[AUT5]{N. T. Zinner}
\thanks[coraut]{Corresponding Author, raph@ornl.gov}
\address[AUT1]{Departement f\"ur Physik \& Astronomie, Universit\"at Basel CH-4056 Basel, Switzerland}  
\address[AUT2]{Physics Division, Oak Ridge National Laboratory, Oak Ridge, TN 37831-6354, USA}
\address[AUT3]{Gesellschaft f\"ur Schwerionenforschung, D-64291 Darmstadt, Germany}
\address[AUT4]{Departament de F{\' i}sica i Enginyeria Nuclear, Universitat Polit\`ecnica de Catalunya, E-08034 Barcelona, Spain}
\address[AUT5]{Institute for Physics and Astronomy, University of \AA rhus, DK-8000 \AA rhus C, Denmark}
\begin{abstract}
Core collapse supernovae are the leading actor in the story of the cosmic origin of the chemical elements.  Existing models, which generally assume spherical symmetry and parameterize the explosion, have been able to broadly replicate the observed elemental pattern. However, inclusion of neutrino interactions produces noticeable improvement in the composition of the ejecta when compared to observations.  Neutrino interactions may also provide a supernova source for light p-process nuclei.  
\end{abstract}

\end{frontmatter}

\section{Modeling Core Collapse Supernova Nucleosynthesis}\label{sect:current}

The complexity of neutrino transport and the frequent failure of self-consistent models for core collapse supernovae to produce explosions have generally divorced modeling of core collapse supernova nucleosynthesis from modeling of the central engine.  Nucleosynthesis simulations commonly replace the central engine of the supernova with a parameterized kinetic energy \emph{piston} \citep{WoWe95,RHHW02,LiCh03} or a thermal energy \emph{bomb} \citep{ThNH96,NaSS98}.  The energy of this blast wave, together with the placement of the \emph{mass cut} that demarcates ejecta from matter that is assumed to fall back onto the neutron star, are tuned to recover the desired explosion energy and ejected \nuc{56}{Ni} mass.   These two methods are largely compatible with the largest differences coming in the inner regions of the ejecta \citep{AuBT91}.  It is this inner region, where much of the iron, nickel and neighboring nuclei are produced, that is also most affected by the details of the explosion mechanism, including the effects of interactions between nuclei and the tremendous neutrino flux.

While the importance of neutrino interactions is manifest in the name of the $\nu$\emph{-process} and well documented for the r-process, neutrinos potentially impact all stages of supernova nucleosynthesis.  During explosive nucleosynthesis, in the inner layers of the ejecta, where iron group nuclei result from $\alpha$-rich freezeout, interactions with neutrinos alter the neutronization, changing the ultimate composition.  
Galactic chemical evolution calculations and the relative neutron-poverty of terrestrial iron and neighboring elements place strong limits on the amount of neutronized material that may be ejected into the interstellar medium by core collapse supernovae \citep{Trim91}.  \citet{HWFM96} placed a limit of $10^{-4} \msun$ on the typical amount of neutron-rich $(\ye\lesssim0.47)$ ejecta allowed from a core collapse supernova.  Those multi-dimensional simulations of the central engine that produce explosions, \citep[see, \eg,][]{HBHF94,JaMu96} predict the ejection of much larger quantities of neutron-rich iron group elements than this limit.  In an effort to compensate, modelers have been forced to invoke the fallback of a considerable amount of matter onto the neutron star, occurring on a timescale longer than was simulated.  One common property exhibited by recent multi-group simulations \citep{LMTM01,RaJa02,ThBP03,BRJK03} is a decrease in the neutronization of the inner layers of the ejecta due to these neutrino interactions.  This is a feature that current parameterized nucleosynthesis models can not replicate because they ignore the neutrino transport.  While the decreased neutronization seen in multi-group transport models would reduce the need to invoke fallback, it also makes any fallback scenario more complicated, since the most neutron-rich material may no longer be the innermost.  

\section{The Nucleosynthesis Implications of Neutrino Interactions}\label{sect:nu}

Because of the impact of the neutrinos on the nucleosynthesis, the nucleosynthesis products from future explosion simulations (utilizing multi-group neutrino transport) will be qualitatively different from parameterized bomb or piston nucleosynthesis models.  This was demonstrated by exploratory calculations by \citet{McFW96}.    The dominant processes are $\nu/\bar{\nu}$ and $\rm e^{\pm}$ captures on shock dissociated free nucleons, though at later times (and in regions with cooler peak temperatures) the more poorly known $\nu/\bar{\nu}$ and $\rm e^{\pm}$ captures on heavy nuclei may contribute significantly.  In addition to their impact on the electron fraction, these interactions, as well as neutral current inelastic neutrino scattering off these nuclei \citep{BrHa91}, are also important to the thermal balance, potentially affecting the $\alpha$-richness of the ejecta, thereby altering the abundance of important nuclei like \nuc{44}{Ti}, \nuc{57}{Fe}, \nuc{58}{Ni} and \nuc{60}{Zn} \citep{WoWe95}.   \citet{TrKM03} used a parameterized neutrino luminosity and spectrum \citep{JaMu96} to drive a supernova explosion and a tracer particle approach with a large nuclear network to examine the nucleosynthesis.   They found significant impact of nuclear electron capture since some ejected zones reached densities $> 10^8 \gcc$.  However, since these authors ignored neutrino captures, their simulations tell only half of the story.  

We have examined the effects of both electron and neutrino captures in the context of recent multi-group supernova simulations.  These models \citep[see][for more details]{FHLM05} are based on fully general relativistic, spherically symmetric simulations \citep{LMTM01}.  \cite{PWBJ05} have performed similar simulations using tracer particles from two dimensional simulations \citep{BRJK03}. In both cases, artificial adjustments to the simulations were needed to remedy the failure of the underlying models of the central engine to produce explosions.  Also in both cases, the simulations were mapped to simplified models as later times, because the neutrino tranport simulations could not be run to sufficiently late times.  While both of these shortcomings need to be addressed, these simulations nonetheless reveal the significant impact that neutrino interactions have on the composition of the ejecta.  

We observe three distinct phases in the evolution of the electron fraction of the matter that will become the innermost ejecta as it collapses, passes through the stalled shock and is driven off by neutrino heating.  During core collapse, the electron fraction in these lower density, silicon-rich, regions is little changed either by the electron capture that is deleptonizing denser regions or the relatively weak neutrino flux.   However, the combination of the larger neutrino flux after core bounce and the burning of silicon to iron in the still infalling matter greatly enhances the neutrino capture rates.  With most of the matter still tied up in relatively inert heavy nuclei, the greater abundance of free protons over free neutrons allows antineutrino captures to dominate, lowering \ye. The passage of the matter through the stalled shock raises the temperature, dissociating the nuclei, but the concomitant increase in density prevents the lifting of the electron degeneracy.  As a result of the high electron chemical potential, the balance of electron and positron captures strongly favors lower electron fraction.  However, the combination of neutrino and antineutrino captures favors higher \ye\ because of the slight dominance of neutrinos over antineutrinos as well as slightly higher abundance of neutrons compared to protons in the fully dissociated, mildly neutron-rich matter.  As a result, the electron fraction undergoes only mild excursions in this phase.  Eventually, continued neutrino heating (or perhaps some other mechanism) is sufficient to reenergize the shock, in the process lifting the electron degeneracy in this innermost ejecta.  As a result, the rate of electron captures drops while the rate of positron captures increases causing \ye\ to rise.   While the dominance of neutrino captures over antineutrino captures drops as the matter becomes neutron-poor, their sum continues to favor higher \ye.   Eventually, the electron chemical potential drops below half the mass difference between the neutron and proton, allowing positron and neutrino captures to dominate electron and antineutrino captures \citep{Belo03}.  With both neutrino emission and absorption processes favoring a higher electron fraction, \ye\ rises markedly in this phase, reaching values as high as 0.55.  

\begin{figure}[tb]
\begin{center}
\includegraphics[width=.9\columnwidth]{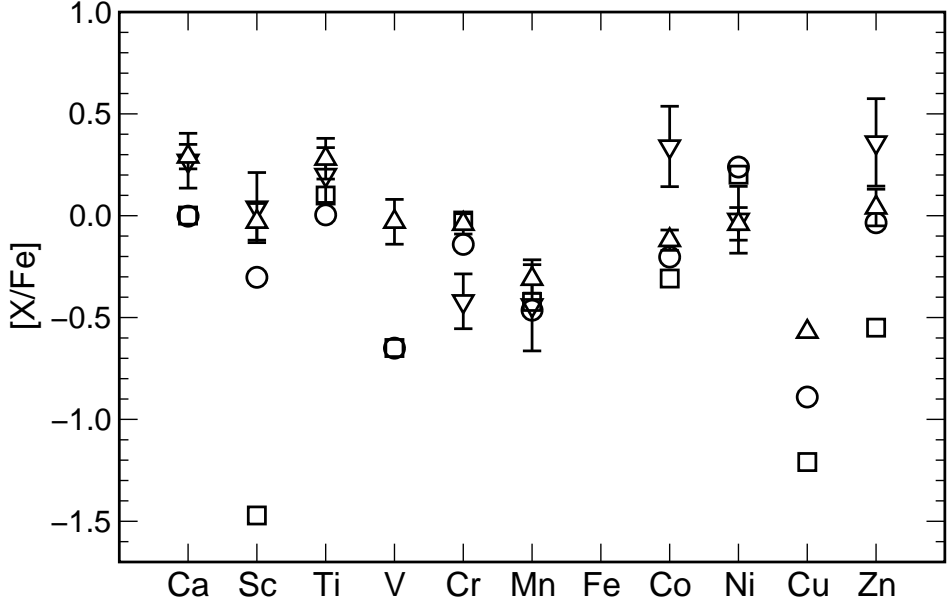}
\caption{Comparison of elemental abundances for Ca to Zn between models by \citet{FHLM05} (circles) and \citet{ThNH96} (squares) and observation determinations for metal-poor \citep{GrSn91} (upward pointing triangles) and extremely metal-poor \citep{CDSH04} (downward pointing triangle) stars.}
\label{fig:nu_elem}
\end{center}
\end{figure}

The global effect of this proton-rich ejecta is the replacement of previously documented overabundances of neutron rich iron peak nuclei (near the N=50 closed shell) \citep{WoWe95,ThNH96} with a mix of \nuc{56}{Ni} and \alp-particles. 
Production of \nuc{58,62}{Ni} is suppressed while \nuc{45}{Sc} and \nuc{49}{Ti} are enhanced.  Elemental abundances of scandium, cobalt, copper and zinc are significantly closer to those observed (see Fig.~\ref{fig:nu_elem}).  The results are however sensitive to the details of the simulations.  \citeauthor{PWBJ05} found a significant sensitivity in the nuclear production to the expansion rate of the matter, which was a parameter in their late time extrapolation.  In addition to the global effects on the neutronization and entropy of the matter, our simulations, which include neutrino and antineutrino capture rates on heavy nuclei \citep{ZiLa05}, find that these reactions have direct impact on the abundances of species like \nuc{53,54}{Fe}, \nuc{55,56,57}{Co}, \nuc{59}{Ni} and \nuc{59}{Cu} at the 10-20\% level.  

\begin{figure}
\begin{center}
\includegraphics[width=\linewidth]{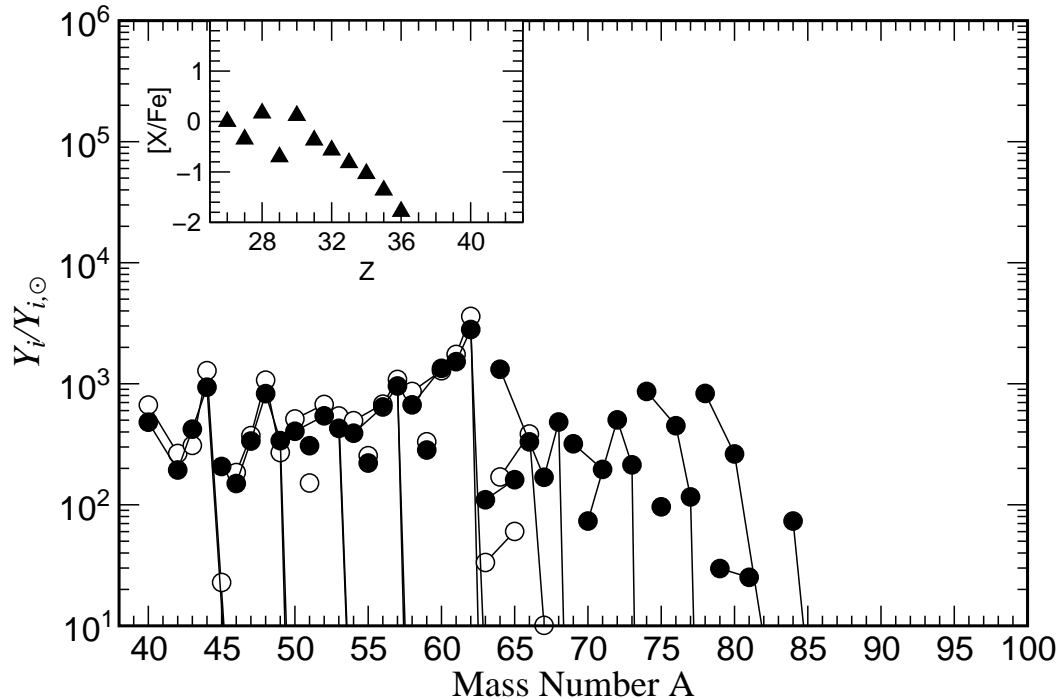}\\
\caption{Isotopic abundances including the effects of neutrino interactions\citep{FMLT05} relative to solar abundances (filled circles) compared with earlier predictions \citep{ThNH96} (open circles), which neglected neutrino interactions.  The effect of neutrino interactions is clearly seen for nuclei above $A>64$ where enhanced abundances are obtained. The inner panel presents the elemental abundances obtained from the isotopic abundances.\label{fig:nup}}
\end{center}
\end{figure}

A mild rp-process is also seen in this proton-rich ejecta (see Fig.~\ref{fig:nup}).  \citeauthor{PWBJ05}, who neglected neutrino captures in their network, find this effect limited to zinc (A=64) by $\beta$-decay lifetimes of waiting-point nuclei that are longer than the expansion timescale.  We have found that transformation of protons into neutrons by neutrino captures allows (n,p) reactions to take the place of $\beta$-decays allowing significant flow to much higher A.  We term this the $\nu$p process because of the essential role the neutrinos play in producing these light p-process nuclei.  The quantity of p-process nuclei produced and upper mass limit of this production are quite  sensitive to the strength of the neutrino interactions and therefore the details  of the neutrino source as well as the proximity and duration of the neutrino exposure.

\section{Conclusion}
Our results, and those of \cite{PWBJ05}, clearly illustrate the need to include the full effect of the supernova neutrino flux on the nucleosynthesis if we are to accurately calculate the iron-peak nucleosynthesis from core collapse supernovae.  The sensitivities displayed in these models point strongly to the need to couple simulations of core collapse supernova nucleosynthesis to models of the explosion mechanism.  Not only will this foster better understand the contribution of supernovae to our cosmic origins, but comparison of nucleosynthesis estimates with observations will also improve our understanding of the central engine.  

\ack The authors wish to thank H.-Th. Janka and A. Mezzacappa for fruitful discussions.  The work has been partly supported by the U.S. National Science Foundation under contract PHY-0244783, by the U.S. Department of Energy, through the Scientic Discovery through Advanced Computing Program, by the Swiss SNF grant 200020-105328 and by the Spanish MCyT and European Union ERDF under contracts AYA2002-04094-C03-02 and AYA200306128.  Oak Ridge National Laboratory is managed by UT-Battelle, LLC, for the U.S. Department of Energy under contract DE-AC05-00OR22725.


\end{document}